\documentclass[preprint,aps,amsmath,nofootinbib,tightenlines,superscriptaddress]{revtex4}
\usepackage{graphicx}
\usepackage{bm}
\usepackage{epsfig}

\newif\ifpdf
\ifx\pdfoutput\undefined
\pdffalse 
\else
\pdfoutput=1 
\pdftrue
\fi

\def\Bslash{B\!\!\!\!\slash}
\def\Dslash{D\!\!\!\!\slash}

\def\Mslash{M\!\!\!\!\!\slash}
\def\nslash{n\!\!\!\slash}
\def\bnslash{\bar n\!\!\!\slash}

\def\OMIT#1{}

\newcommand{\nn}{\nonumber} 

\newcommand{\bn}{{\bar n}}
\newcommand{\bnP}{\bar {\cal P}}
\newcommand{\mcdot}{\!\cdot\!}

\newcommand{\cD}{{\cal D}}

\newcommand{\cP}{{\cal P}}

\newcommand{\bea}{\begin{eqnarray}}
\newcommand{\eea}{\end{eqnarray}}

\newcommand{\SCETa}{SCET$_{\rm I}$ }
\newcommand{\SCETb}{SCET$_{\rm II}$ }

\begin{document}
\ifpdf
\DeclareGraphicsExtensions{.pdf, .jpg}
\else
\DeclareGraphicsExtensions{.eps, .jpg}
\fi


\preprint{ \vbox{\hbox{ } \hbox{hep-ph/0303156} \hbox{MIT-CTP-3349} \hbox{UCSD/PTH 03-01}  
}}

\title{\phantom{x}
\vspace{0.5cm} 
On Power Suppressed Operators and Gauge Invariance in SCET
\vspace{0.6cm} 
}

\author{Christian W.~Bauer}
\affiliation{Department of Physics, University of California at San Diego,\\
        La Jolla, CA 92093\footnote{Electronic address: bauer@physics.ucsd.edu}
        \vspace{0.15cm}}
\author{Dan Pirjol}
\affiliation{Department of Physics and Astronomy, The Johns Hopkins University,
\\
        Baltimore, MD 21218\footnote{Electronic address: dpirjol@pha.jhu.edu}
        \vspace{0.15cm}}
\author{Iain W.~Stewart\vspace{0.6cm}}
\affiliation{Center for Theoretical Physics, 
  Massachusetts Institute of  Technology,
  \\ 
  Cambridge, MA 02139\footnote{Electronic address: iains@mit.edu}
  \vspace{0.6cm}
}

\vspace{0.3cm}

\begin{abstract}
\vspace{0.3cm}
\setlength\baselineskip{18pt}

The form of collinear gauge invariance for power suppressed operators in the
soft-collinear effective theory is discussed.  Using a field redefinition we
show that it is possible to make any power suppressed ultrasoft-collinear
operators invariant under the original leading order gauge transformations. Our
manipulations avoid gauge fixing.  The Lagrangians to ${\cal O}(\lambda^2)$ are
given in terms of these new fields.  We then give a simple procedure for
constructing power suppressed soft-collinear operators in \SCETb by using an
intermediate theory \SCETa.

\end{abstract}

\maketitle




\section{Introduction}

The soft-collinear effective theory (SCET) has been proposed as a systematic
approach for separating hard and soft scales in processes with energetic quarks
and gluons~\cite{bfl,bfps,cbis,bps2}.  The infrared physics is described in the
effective theory in terms of collinear, soft, and ultrasoft fields with well
defined momentum scaling.  These fields are used to construct operators such as
Lagrangians and currents that describe long distance effects, while hard
corrections are contained in Wilson coefficients. This formalism builds on and
extends earlier techniques used for discussing factorization~\cite{reviews}.

The degrees of freedom in SCET include collinear quarks $\xi_n$ and gluons
$A_n^\mu$ with momentum scaling $p_c^\mu = (n\cdot p, \bn\cdot p, p_\perp) \sim
Q(\lambda^2, 1, \lambda)$, soft modes $q_s, A_s^\mu$ with momenta $p_s^\mu \sim
Q\lambda$, and ultrasoft (usoft) modes $q_{us}, A_{us}^\mu$ with momenta $p_{\rm
  us}^\mu \sim Q\lambda^2$.  Here $Q$ is the hard scale, $\lambda\ll 1$ is the
expansion parameter, and $n_\mu, \bn_\mu$ are two light-cone unit vectors
satisfying $n^2 = \bn^2 = 0$ and $n\cdot \bn = 2$. The explicit set of required
fields may differ depending on the relevant scales in a given process. For
instance, in Drell-Yan it is useful to have collinear fields for two light-like
directions and for multijet-production more than two directions are
required~\cite{bfprs,jets}.

In many exclusive heavy meson decays to energetic light hadrons there are
important effects at the scales $Q^2$, $Q\Lambda$, and $\Lambda^2$, where
$\Lambda\sim 0.5\,{\rm GeV}$ is a hadronic scale. To correctly account for these
effects, a sequence of two effective theories, SCET$_{\rm I}$ and SCET$_{\rm
  II}$, can be used~\cite{bps4}.\footnote{In Ref.~\cite{bps2} a version of SCET
  was constructed that simultaneously involves collinear, soft, and usoft
  fields. While it is possible that some physical process may simultaneously
  require these degrees of freedom, here we restrict ourselves to the
  degrees of freedom of \SCETa-\SCETb which suffice for most applications.}
One thus distinguishes between
\begin{eqnarray} \label{SCET12}
 && \mbox{ SCET$_{\rm I}$}: \qquad\  \mbox{collinear fields with
  $(p^+_c,p^-_c,p^\perp_c)\sim Q(\lambda^2,1,\lambda)$ and usoft fields}\nn\\
 && \mbox{\hspace{2.6cm}  with
  $p^\mu_{us} \sim Q\lambda^2$ where $\lambda \sim \sqrt{\Lambda/Q}$} \nn \\
 && \mbox{ SCET$_{\rm II}$}: \qquad \mbox{collinear fields with
  $(p^+_c,p^-_c,p^\perp_c)\sim Q(\eta^2,1,\eta)$ and soft fields} \nn\\
 &&\mbox{\hspace{2.6cm}  with $p^\mu_s \sim Q\eta$ where
  $\eta \sim \Lambda/Q$ \,. } \nn
\end{eqnarray}
For clarity the power counting parameter $\eta$ is used for SCET$_{\rm II}$
rather than $\lambda$.  In exclusive processes the energetic/soft hadrons are
described by collinear/soft fields in SCET$_{\rm II}$. Both fields have
$p_\perp\sim \Lambda$ which is appropriate for describing the constituents of
hadrons of size $r_\perp \sim 1/\Lambda$.  For exclusive processes the theory
SCET$_{\rm I}$ plays an intermediate role by describing in a local way the
fluctuations with $p^2\sim Q\Lambda$ that are involved in interactions between
soft and collinear fields in SCET$_{\rm II}$.  In contrast, SCET$_{\rm I}$
suffices for describing factorization in inclusive processes like $B\to
X_s\gamma$, as well as some exclusive processes like $B\to \gamma
e\nu$~\cite{bps2,bgamenu}. Interactions in SCET$_{\rm II}$ are discussed
in Refs.~\cite{bps2,HN} and power corrections in SCET$_{\rm I}$ were studied in
Refs.~\cite{chay,chay2,bpspc,bcdf,bps4,ps1,bf2,ira}. Quark masses were
considered in Ref.~\cite{adam}.

The symmetries of the effective theory provide an important guiding principle
for constraining the form of operators, especially at the level of power
corrections. The SCET has a rich symmetry structure, reflecting the interplay
between the different length scales it describes. The constraints include power
counting, collinear/soft/ultrasoft gauge invariance, reductions in spin
structures, and a reparameterization
invariance~\cite{bfl,bfps,cbis,bps2,chay,rpi,bpspc} (see Ref.~\cite{rev2} for a
brief review of the symmetries).  At a given order in $\lambda$ the most general
set of operators for a given process can be constructed using
\begin{itemize}
\item {\it Power counting:}\quad 
  Restricts the type of fields and derivatives allowed in the operator
\item {\it Gauge invariance:}\quad 
  Requires operators to be built out of gauge invariant building blocks. 
\item {\it Reparameterization invariance:}\quad 
  Corresponds to the restoration of Lorentz invariance order by order in
  $\lambda$. 
\item {\it Locality:}\quad 
    The theory \SCETa is only non-local in ${\cal O}(Q)$ momenta. Only inverse
    powers of the large label momentum are allowed and collinear Wilson lines
    have to be built out of ${\cal O}(1)$ gluons. 
\end{itemize}
Note that \SCETa is constructed in a local manner, but after doing this it is
useful to consider a field redefinition $\xi_n \to Y\xi_n$ which introduces
non-locality at the usoft scale. The locality restriction does not apply to
\SCETb. Integrating out $p^2\sim Q\Lambda$ modes immediately results in
operators involving the soft Wilson line $S$~\cite{bps2}, and it contains
inverse powers of $1/\Lambda$ momenta.  In the following we will focus on gauge
invariance and discuss subtleties which arise in constructing invariant
operators at subleading order.

\begin{table}[t!]
\begin{center}
\begin{tabular}{cc|cccc}
  & \hspace{0.6cm}Object \hspace{0.6cm}  & Collinear ${\cal U}_c$
    & Usoft $U_{us}$ & \\ \hline
  & $\xi_{n}$ & ${\cal U}_c\ \xi_{n}$ &   $U_{us}\,\xi_{n}$ \\
  & $gA_{n}^\mu$ & \hspace{0.2cm}${\cal U}_c\: gA^\mu_{n}\: {\cal U}_c^{\dagger} +
    {\cal U}_c \big[i\cD^\mu , {\cal U}_c^{\dagger} \big]$ &
    $U_{us}\, gA_{n}^\mu\, U_{us}^{\dagger}$ \\
  & $W$ & ${\cal U}_c\, W$ & $U_{us}\, W\, U_{us}^{\dagger}$ & \\[3pt]
 \hline
  & $q_{us}$ & $q_{us}$ & $U_{us}\, q_{us}$  & \\
  & $gA_{us}^\mu$ & $gA_{us}^\mu$ & $U_{us}  gA_{us}^\mu U^\dagger_{us} +
   U_{us} [i\partial^\mu , U_{us}^{\dagger}]$  & \\
    & $Y$ & $Y$  & $U_{us}\, Y$ & \\
 \hline
\end{tabular}
\end{center}
{\caption{\setlength\baselineskip{12pt}
Gauge transformations for the collinear and usoft
fields from Ref.~\cite{bps2}, where $i\cD^\mu \equiv \frac{n^\mu}{2}\,
\bnP + {\cal P}_\perp^\mu +  \frac{\bn^\mu}{2}\, i\, n\mcdot D_{\rm us} $. The
collinear fields and transformations are understood to have momentum labels and
involve convolutions, but for simplicity these indices are suppressed. The usoft
transformations do not change the momentum labels of collinear fields.  }
\label{table_gt} }
\end{table}

\begin{table}[t!]
\begin{center}
\begin{tabular}{cc|cccc}
   & \hspace{0.6cm}Objects\hspace{0.6cm}  & Collinear ${\cal U}_c$ & Soft $U_s$
     \\ \hline
  & $\xi_{n}$ & ${\cal U}_c\ \xi_{n}$ &   $\xi_{n}$ \\
  & $gA_{n}^\mu$ & \hspace{0.2cm}${\cal U}_c\: gA^\mu_{n}\: {\cal U}_c^{\dagger} +
    {\cal U}_c \big[ i\partial_c^\mu {\cal U}_c^{\dagger} \big]$ &
    $ gA_{n}^\mu $ \\
  & $W$ & ${\cal U}_c\, W$ & $W$ & \\[3pt]
 \hline
  & $q_{s}$ & $q_{s}$  & $U_s\, q_{s}$  \\
  & $g A_{s}^\mu$ & $g A_{s}^\mu$ &  $U_{s}\, g A_{s}^\mu U_s^{\dagger} +
    U_s [ i \partial_s^\mu, U_s^{\dagger}]$  \\
   & $S$ & $S$ & $U_s\, S$ \\
\end{tabular}
\end{center}
{\caption{\setlength\baselineskip{12pt}Gauge transformations for collinear
and soft fields in ${\rm SCET}_{\rm II}$ from Ref.~\cite{bps2}. Momentum labels
are suppressed, and $\partial_c^\mu$ and $\partial_s^\mu$ are defined to only
pick out collinear and soft momenta respectively.
Here $i\partial_c^\mu\ne i {\cal D}^\mu$ since usoft fields are not included in
${\rm SCET}_{\rm II}$.}
\label{table_gt2} }
\end{table}

The gauge transformations for the SCET fields were derived in \cite{bps2} and
are summarized in Tables~\ref{table_gt} and \ref{table_gt2}. Here
$\partial_c^\mu {\cal U}_c \sim Q(\lambda^2,1,\lambda)$, $\partial_s^\mu U_s
\sim Q\lambda$, and $\partial^\mu U_{us} \sim Q\lambda^2$ distinguish the
collinear, soft and usoft gauge transformations respectively. Partial
derivatives without a subscript are usoft, so $i\partial_\mu\sim Q\lambda^2$. In
Table I we have used
\begin{eqnarray} \label{Dc}
  i\cD^\mu \equiv \frac{n^\mu}{2}\, \bnP + {\cal P}_\perp^\mu +
   \frac{\bn^\mu}{2}\, i\, n\mcdot D_{\rm us}
\end{eqnarray}
in the fundamental representation. Note that only the $n \mcdot A_{us}$
component of the usoft gauge field appears here and that the components of
$\cD^\mu$ have the same scaling in $\lambda$ as the collinear gluon field, so
all transformations are homogeneous.  Thus, power counting strongly constrains
the leading usoft-collinear interactions. It also forces us to have a multipole
expansion so that only the $n\mcdot k$ momenta of collinear particles can be
changed by interactions with usoft gluons. In Refs.~\cite{bfl,bfps,cbis,bps2}
this expansion is done in momentum space while in Refs.~\cite{bcdf,HN,bf2} it is
done in position space. This leads to formulations of SCET whose operators
appear slightly different, but whose final predictions for physical observables
have to be the same.

In this paper we discuss how gauge invariance is realized for power suppressed
operators in both ${\rm SCET}_{\rm I}$ and ${\rm SCET}_{\rm II}$. ${\rm
  SCET}_{\rm I}$ is studied in Section \ref{sect1} where we clarify the nature
of collinear gauge invariance in power suppressed operators with ultrasoft
derivatives. This is done by showing that it is possible to arrange these power
suppressed operators such that only the original {\em leading order} gauge
transformations are needed at any order in the power expansion.  This was also
the goal of a recent study by Beneke and Feldmann~\cite{bf2} and a comparison is
given with their results.  The form of our transformed fields is different than
theirs, reflecting a freedom in choice of viable field redefinitions.  We found
that it was not necessary to do any gauge fixing in our manipulations.

In SCET$_{\rm II}$ the soft and collinear gauge invariance alone allow a large
number of operators, reflecting the more non-local nature of this theory. In
particular, gauge invariance does not uniquely fix the path of the Wilson lines.
However, since SCET$_{\rm II}$ is matched on from SCET$_{\rm I}$ and not from
full QCD, one can obtain information about the operators relevant for a given
process from the structure of operators in SCET$_{\rm I}$. We illustrate the
SCET$_{\rm I}$\,$\to$ SCET$_{\rm II}$ matching by several examples in
Section~\ref{sect2}.

\section{Gauge Invariance in SCET$_{\rm I}$}\label{sect1}

At leading order the SCET$_{\rm I}$ Lagrangian for collinear
quarks is~\cite{bfps,cbis}
\begin{eqnarray}\label{L0}
{\cal L}_{\xi\xi}^{(0)} 
&=& \bar \xi_n \left[ i n\mcdot D
+ i \Dslash^\perp_c  W \frac{1}{\bnP} W^\dagger 
  i \Dslash^\perp_c\right] \frac{\bnslash}{2} \xi_n  \,,
\end{eqnarray}
where the collinear covariant derivatives are $iD_c^\mu = {\cal P}^\mu +
gA_n^\mu$ with label operators $\cP^\mu$, the full derivative $in\mcdot
D=in\mcdot\partial + g n\mcdot A_{us} + gn\mcdot A_n$, and the Wilson line $W$
is built out of $\bn\mcdot A_n$ fields where $f(i\bn\mcdot D_c) = W f(\bnP)
W^\dagger$
\begin{eqnarray} \label{W}
   W =  \Big[ \sum_{\rm perms} \exp\Big( -\frac{g}{\bnP}\: 
  \bn\mcdot  A_{n,q}(x) \ \Big) \Big] \,.
\end{eqnarray}
Under the gauge transformations in Table~\ref{table_gt} covariant derivatives
acting in the fundamental representation transform under collinear and usoft
transformations as
\begin{eqnarray}\label{Dtrans1}
{\cal U}_c &:&  \quad in\mcdot D \to {\cal U}_c\, in\mcdot D \, {\cal
  U}_c^{\dagger}\,, 
 \phantom{{}_c}\qquad 
  iD_c^\perp \to {\cal U}_c\, i D_c^\perp {\cal U}_c^{\dagger}\,,\qquad
  i\bn\mcdot D_c \to {\cal U}_c\, i\bn\mcdot D_c \, {\cal U}_c^{\dagger}\,, \\
 U_{us} &:&    \quad in\mcdot D \to U_{us}\, in\mcdot D \, U_{us}^{\dagger}\,,\qquad 
  iD_c^\perp \to U_{us}\, i D_c^\perp U_{us}^{\dagger}\,,\qquad
  i\bn\mcdot D_c \to U_{us}\, i\bn\mcdot D_c \, U_{us}^{\dagger}\,.\nn
\end{eqnarray}
It is straightforward to verify that all factors of ${\cal U}_c$ or $U_{us}$ drop
out of ${\cal L}_{\xi\xi}^{(0)}$, which has been shown to be the most general
possible operator consistent with gauge invariance, power counting, and
reparameterization invariance~\cite{bps2,rpi}.  The same is true of the leading
order collinear gluon action
\begin{eqnarray} \label{Lcg}
  {\cal L}_{cg}^{(0)} &=&  \frac{1}{2 g^2}\, {\rm tr}\ \bigg\{ 
    \Big[i\cD^\mu +g A_{n,q}^\mu \,, i\cD^\nu + g A_{n,q'}^\nu \Big] \bigg\}^2 \\
 && +2\, {\rm tr} \Big\{ \, \bar c_{n,p'}\,  \Big[ i\cD_\mu , \Big[ i\cD^\mu + 
     g A_{n,q}^\mu \,, c_{n,p}\,\Big]\Big] \Big\}   
   + \frac{1}{\alpha}\, {\rm tr}\ \Big\{ [i\cD_\mu\,, A_{n,q}^\mu]\Big\}^2\,. \nn
\end{eqnarray}
The terms on the second line are the gauge fixing terms for a general
covariant gauge, where $c_n$ are adjoint ghost fields.

Beyond leading order the form of the subleading Lagrangians can be determined by
matching calculations and use of the SCET symmetries.  There is a
reparameterization invariance~\cite{LM} (RPI), which in SCET is due to the
freedom in choosing the basis vectors $n$ and $\bn$, and in decomposing the
momenta $\bn\mcdot(p+k)$ and $(p_\perp^\mu+k_\perp^\mu)$ into collinear $p$ and
usoft $k$ components~\cite{chay,rpi}.  This RPI connects collinear and usoft
derivatives,
\begin{eqnarray} \label{rpid}
 \bnP+ i\bn\mcdot\partial\,,\qquad \cP_\perp^\mu+i\partial_\perp^\mu\,,
\end{eqnarray}
and also relates the Wilson coefficients of leading and subleading
operators~\cite{chay,rpi,bcdf,ps1}. 

To turn the derivatives in Eq.~(\ref{rpid}) into covariant derivatives we make
use of gauge symmetry. This forces the label operator to be replaced by the
collinear covariant derivative $iD_c^\mu$, but as we shall see it allows some
freedom in the usoft term~\cite{chay2}.  In Refs.~\cite{chay,rpi} the usoft
derivative was made covariant with the choice $iD_{us}^\mu$, so the RPI
combinations in Eq.~(\ref{rpid}) become
\begin{eqnarray} \label{c1}
 && \mbox{choice i)}\qquad\qquad 
   i\bn\mcdot D= i\bn\mcdot D_c + i\bn\mcdot D_{us}\,,\qquad
   i D_\perp^\mu = i D_{c,\perp}^\mu + iD_{us,\perp}^\mu  \,.\qquad\qquad\qquad
\end{eqnarray}
For the purpose of gauge transformations this corresponds to promoting the
ultrasoft field to a full background field of a quantum collinear gauge field
so that
\begin{eqnarray}\label{gauge1}
 gA_n^\mu \to {\cal U}_c\, gA_n^\mu \, {\cal U}_c^{\dagger} 
  + {\cal U}_c [{\cal P}^\mu + iD^\mu_{\rm us}\,, {\cal U}_c^{\dagger}]\,,
\end{eqnarray}
and the combined field $A^\mu = A_n^\mu + A_{\rm us}^\mu$ transforms as
\begin{eqnarray}
  gA^\mu \to {\cal U}_c gA^\mu {\cal U}_c^{\dagger} + {\cal U}_c [{\cal P}^\mu 
    + i\partial_{\rm us}^\mu\,, {\cal U}_c^{\dagger}] \,.
\end{eqnarray}
With this choice one still has homogeneous gauge transformations in
Table~\ref{table_gt} at leading order, which we will call $G^{(0)}$, however one
also induces subleading collinear transformations for $A_n^\perp$ and
$\bn\mcdot A_n$ suppressed by $\lambda$ and $\lambda^2$ respectively
\begin{eqnarray}
  G^{(1)}:\qquad  A^\mu_{n,\perp} \to {\cal U}_c [iD_{\perp,us}^\mu , 
  {\cal U}_c^\dagger ] \,,
  \qquad \bn\mcdot A_n \to {\cal U}_c [i\bn\mcdot D_{us} , {\cal
    U}_c^\dagger ] \,.
\end{eqnarray}
Thus, much like the reparameterization invariance, there are gauge
transformations that connect the leading and subleading terms. This observation
was first made in Ref.~\cite{chay2}.  For example, using the gauge completion
given in Eq.~(\ref{c1}) the $O(\lambda)$ Lagrangian is
\begin{eqnarray}\label{L1}
{\cal L}_{\xi\xi}^{(1)}  &=& \bar \xi_n \Big[ i \Dslash^\perp_{\rm us} 
 \frac{1}{\bn\mcdot iD_c}  i \Dslash^\perp_c 
  + i \Dslash^{\perp}_c \frac{1}{\bn\mcdot iD_c} i \Dslash^\perp_{us}\Big] 
  \frac{\bnslash}{2} \xi_n\,.
\end{eqnarray}
Under a collinear gauge transformation $G^{(0)}$ from Table~\ref{table_gt} one
finds
\begin{eqnarray} 
  {\cal L}^{(1)} \to {\cal L}^{(1)} - \bar \xi_n \left[ 
   [i\Dslash^\perp_{\rm us},  {\cal U}_c^\dagger] {\cal U}_c 
   \frac{1}{\bn\mcdot iD_c} 
    i \Dslash^\perp_c + i \Dslash^{\perp}_c \frac{1}{\bn\mcdot iD_c} 
   {\cal U}_c^\dagger [i\Dslash^\perp_{us}, {\cal U}_c] \right]
 \frac{\bnslash}{2} \xi_n\,.  
\end{eqnarray} 
The second term cancels against the $G^{(1)}$ variation of the leading order
Lagrangian ${\cal L}_{\xi\xi}^{(0)}$, implying that the effective Lagrangian is
invariant up to this order. The other subleading actions with usoft fields
are~\cite{bcdf,bps4,feldmann,ps1}
\begin{eqnarray} \label{subLs}
  {\cal L}^{(2a )}_{\xi q} &=& \bar\xi_n  \frac{1}{i\bn\mcdot D_c}\: 
     ig \big\{ \Mslash_\perp + \frac{\bnslash}{2} n\mcdot M \big\}
     \, W \, q_{us} \mbox{ + h.c.} \,, \nn\\
 {\cal L}_{cg}^{(1)} &=& \frac{2}{g^2}\: {\rm tr} 
  \Big\{ \big[i {D}_0^\mu , iD_c^{\perp\nu} \big] 
         \big[i {D}_{0\mu} , iD_{us\,\nu}^\perp \big] \Big\}\,,\\
 {\cal L}_{cg}^{(2)} &=&  \frac{1}{g^2}\: {\rm tr} 
  \Big\{ \big[i {D}_0^\mu , iD_{us}^{\perp\nu} \big] 
         \big[i {D}_{0\mu} , iD_{us\,\nu}^\perp \big] \Big\}
  + \frac{1}{g^2}\: {\rm tr} 
  \Big\{ \big[i D_{us}^{\perp\mu} , iD_{us}^{\perp\nu} \big] 
         \big[i {D}_{c\mu}^\perp , i{D}_{c\nu}^\perp \big] \Big\}\nn\\
 &+&\frac{1}{g^2}\: {\rm tr} 
  \Big\{ \big[i {D}_0^\mu , i n\mcdot D \big] 
         \big[i {D}_{0\mu} , i \bn\mcdot D_{us} \big] \Big\}
  + \frac{1}{g^2}\: {\rm tr} 
  \Big\{ \big[i D_{us}^{\perp\mu} , iD_{c}^{\perp\nu} \big] 
         \big[i {D}_{c\mu}^\perp , i{D}_{us\nu}^\perp \big] \Big\}\nn
  \,,
\end{eqnarray}
where $i g M^\mu = [i\bn\mcdot D_c, i D_{us}^\mu + \bn^\mu gn\mcdot A_n /2 ]$
and $iD_0^\mu = iD_c^\mu + i \bn^\mu n\mcdot D_{us}/2$. (The terms ${\cal
  L}_{\xi q}^{(1)}$ and ${\cal L}_{\xi q}^{(2b)}$ do not depend on ultrasoft
covariant derivatives and are shown below in Eq.~(\ref{Lxiq}).)  Similar
manipulations show that the results in Eq.~(\ref{subLs}) are invariant with
terms canceled by the $G^{(1)}$ transformation of ${\cal L}_{\xi q}^{(1)}$ and
${\cal L}_{cg}^{(0,1)}$.

Although operators with usoft fields are gauge invariant, the presence of
$G^{(1)}$ requires transformations of operators at different powers in $\lambda$
to cancel one another. This is unsatisfactory since constraining operators at
any particular order requires transforming lower order operators. Furthermore
this would mean we would only be able to assign an unambiguous meaning to the
sum of leading and subleading matrix elements. Instead, we would like to use
fields with no $G^{(1)}$ transformation, so that operators are manifestly
invariant under $G^{(0)}$ at each order in $\lambda$.  In other words the terms
at a given order are invariant without needing the transformation of lower order
terms.  To this end, consider the field redefinitions
\begin{eqnarray}\label{redef}
  g \bn\mcdot \hat A_n =  g\bn\mcdot  A_n -
      {\cal W}[i\bn\mcdot D_{us}, {\cal W}^\dagger ]\,,\qquad
  g \hat A_n^\perp =  g A_n^\perp -
      {\cal W}[i D_{us}^\perp, {\cal W}^\dagger ]\,,\qquad
\end{eqnarray}
where $g n\mcdot \hat A_n = gn\mcdot A_n$, and $\hat A_n^\mu$ are new
collinear gluon fields. Here ${\cal W}$ is the product of Wilson lines defined
in Ref.~\cite{bcdf} which in position space is
\begin{eqnarray} \label{cW}
 {\cal W}(x) &=& P \exp\Big( ig \int_{-\infty}^0\!\!\!\!\! ds\, 
  \bn\mcdot (A_n \!+\!A_{us})(\bn s\!+\!x)  \Big) 
  \bigg[ P \exp\Big( i g \int_{-\infty}^0 \!\!\!\!\! ds\, 
  \bn\mcdot A_{us}(\bn s\!+\!x) \Big)\bigg]^\dagger .
\end{eqnarray}
In Eq.~(\ref{cW}) the collinear fields $A_n^\mu(X\!+\!x)$ are the Fourier
transforms of $A_{n,p}^\mu(x)$ with $X$ the conjugate variable to $p$. Under
collinear gauge transformations ${\cal W}\to U_c {\cal W}$, while under usoft
gauge transformations ${\cal W}\to U_{us} {\cal W} U_{us}^\dagger$.  The
presence of ${\cal W}$ in Eq.~(\ref{redef}) causes $\hat A_n$ to be defined in
terms of a non-linear function of $A_n$.  Note that our transformation in
Eq.~(\ref{redef}) differs from that in Ref.~\cite{bf2}, as we discuss in more
detail below. Under a collinear gauge transformation the $\perp$ component of
the new collinear gluon field transforms as (suppressing momentum space labels)
\begin{eqnarray}
  g \hat A^\perp_n &\to & {\cal U}_c\, g A_n^\perp {\cal U}_c^\dagger 
   +  {\cal U}_c [\cP_\perp + i D_{us}^\perp , {\cal U}_c^\dagger ] 
   - {\cal U}_c {\cal W} [ i D^\perp_{us}, {\cal W}^\dagger {\cal U}_c^\dagger]
   \nonumber\\
&=& {\cal U}_c\, g A_n^\perp {\cal U}_c^\dagger + {\cal U}_c \cP^\perp 
 {\cal U}_c^\dagger 
  + {\cal U}_c\, iD^\perp_{us} {\cal U}_c^\dagger 
  - {\cal U}_c\, {\cal W} iD^\perp_{us} {\cal W}^\dagger {\cal U}_c^\dagger \nn\\
 &=& {\cal U}_c g\hat A_n^\perp {\cal U}_c^\dagger + {\cal U}_c \cP_\perp 
  {\cal U}_c^\dagger \,.
\end{eqnarray}
Only hatted fields appear in the final result.  With a similar set of steps we
find $g \bn\mcdot\hat A_n \to {\cal U}_c\, g \bn\mcdot\hat A_n {\cal
  U}_c^\dagger + {\cal U}_c \bnP\, {\cal U}_c^\dagger$. Therefore
\begin{eqnarray}\label{gthat} 
  g\hat A_n^\mu \to {\cal U}_c\: g\hat A_n^\mu\, {\cal U}_c^\dagger 
   + {\cal U}_c\, [i{\cal D}^\mu\,, {\cal U}_c^\dagger]\,,
\end{eqnarray}
just like in Table~\ref{table_gt}. Thus, in terms of the hatted fields,
transformations that involve suppressed terms like $G^{(1)}$ never appear. This
is the desired result. 

To express the Lagrangians in terms of hatted fields it is useful to have the
inverse transformation to Eq.~(\ref{redef}).  This is complicated by the factors
of ${\cal W}={\cal W}[\bn\mcdot A_n, \bn\mcdot A_{us}]$ given in
Eq.~(\ref{redef}), which depend non-linearly on the gluon fields. Now, we know that
\begin{eqnarray}
  i\bn\mcdot D\: {\cal W} 
   = {\cal W} g \bn\mcdot A_{us} \,,
\end{eqnarray}
which implies that in terms of the hatted fields ${\cal W}={\cal W}[\bn\mcdot
\hat A_n, \bn\mcdot A_{us}]$ satisfies the equation
\begin{eqnarray}
&& 0 = ( i \bn\mcdot\hat D_c + {\cal W} i\bn\mcdot D_{us} {\cal W}^\dagger )
  {\cal W} - {\cal  W} g\bn\mcdot A_{us} 
  = i\bn\mcdot \hat D_c\: {\cal  W}\,. 
\end{eqnarray}
However, this equation has a unique solution $\hat W$. Switching to momentum
labels and residual coordinates $x$~\cite{cbis}, this $\hat W$ is just the
standard Wilson line in Eq.~(\ref{W}) expressed in terms of the $\bn\mcdot\hat
A_n$ collinear field (since they are defined by the same equation). This gives
the remarkable result that after the field redefinition we have to all orders in
$\lambda$
\begin{eqnarray}
  {\cal W} = \hat W =  \Big[ \sum_{\rm perms} \exp\Big( -\frac{g}{\bnP}\: 
  \bn\mcdot \hat A_{n,q}(x) \ \Big) \Big] \,,
\end{eqnarray}
which is independent of the usoft gauge field.  Under the gauge
transformations $\hat W\to U_c\hat W$ and $\hat W\to U_{us} \hat W U_{us}^\dagger$
just like we had for $W$.  Thus, the inverse transformation to Eq.~(\ref{redef})
can be written
\begin{eqnarray} \label{redef2}
 g \bn\mcdot  A_n =  g\bn\mcdot \hat A_n +
      \hat W [i\bn\mcdot D_{us}, {\hat W}^\dagger ]\,,\qquad
  g  A_n^\perp =  g \hat A_n^\perp +
      {\hat W}[i D_{us}^\perp, {\hat W}^\dagger ]\,,\qquad
\end{eqnarray}
This corresponds to gauging the RPI combinations in Eq.~(\ref{rpid}) to
\begin{eqnarray} \label{c2}
 && \mbox{choice ii)}\qquad\quad
  i\bn\mcdot \hat D = i\bn\mcdot \hat D_c + {\hat W} i\bn\mcdot D_{us} 
   {\hat W}^\dagger  \,,\qquad 
  i \hat D_\perp^\mu = i \hat D_{c,\perp}^\mu + {\hat W} iD_{us,\perp}^\mu 
   {\hat W}^\dagger  \,,\qquad\qquad
\end{eqnarray}
rather than using choice i) in Eq.~(\ref{c1}).  Under a collinear and usoft
gauge transformations these derivatives transform exactly as in
Eq.~(\ref{Dtrans1})
\begin{eqnarray}\label{Dtrans2}
{\cal U}_c :&&   in\mcdot \hat D \to {\cal U}_c\, in\mcdot \hat D \,
 {\cal U}_c^\dagger\,, \phantom{{}_c}\qquad 
  i\hat D_c^\perp \to {\cal U}_c\, i \hat D_c^\perp {\cal U}_c^\dagger\,,\qquad
  i\bn\mcdot \hat D_c \to {\cal U}_c\, i\bn\mcdot \hat D_c\, {\cal U}_c^\dagger\,,
  \\
U_{us} :&&  in\mcdot \hat D \to U_{us}\, in\mcdot \hat D\, U_{us}^\dagger\,,\qquad 
  i\hat D_c^\perp \to U_{us}\, i \hat D_c^\perp U_{us}^\dagger\,,\qquad
  i\bn\mcdot \hat D_c \to U_{us}\, i\bn\mcdot \hat D_c\, U_{us}^\dagger\,.\nn
\end{eqnarray}

In Ref.~\cite{bf2} transformations were also made with the aim of determining
fields that could be used in power suppressed operators while avoiding gauge
transformations that mix different orders in $\lambda$. Similar to the
construction here their initial fields transform as in Eq.~(\ref{gauge1}) and
the desired final collinear transformations are identical to the form in
Ref.~\cite{bps2}, shown in our Table~\ref{table_gt}.  In Ref.~\cite{bf2} the new
collinear quark and gluon fields were defined as
\begin{eqnarray}\label{BF}
   \xi_n &=& R W_c^\dagger \hat \xi_n \,, \\
  g A_{\perp c} &=& R \Big( W_c^\dagger  i  \hat D_{\perp c} W_c^\dagger  -
    i\partial_c^\perp  \Big) R^\dagger\,,\nonumber\\
  gn\mcdot  A_{c} &=& R \Big( W_c^\dagger  i n\mcdot \hat D W_c  -
    i n\mcdot D_{us}(\bn\mcdot x n/2)\Big)  R^\dagger \,,\nonumber 
\end{eqnarray}
where the fields on the left-hand side are understood to be in a light-like
axial gauge with $\bn\mcdot A_c=1$.  The matrix $R$ is
defined as $R(x) = P\exp( ig \int_C dz_\mu A_{\rm us}^\mu(z))$ with the path $C$
a straight line connecting $\frac12 \bn_\mu n\mcdot x$ to $x$. In
Ref.~\cite{bf2} the collinear fields were constructed entirely in position
space, and a multipole expansion was performed on the usoft fields $\phi_{us}(x)
= \phi_{us}(x_-) + (x_\perp \cdot i\partial_\perp) \phi_{us}(x_-) + \ldots$. The
transformation with the matrix $R$ was then necessary to connect collinear and
usoft fields which are at different space-time points.  After inserting these
fields into the effective Lagrangian, operators involving the matrix $R$ were
expanded using the Fock-Schwinger gauge for the ultrasoft gluon field.

The results in Eq.~(\ref{BF}) differ from our field transformation in
Eq.~(\ref{redef2}) in several respects. First, we did not need to redefine the
collinear quark field $\xi_{n,p}(x)$ since our labeled collinear fields carry
residual ultrasoft momentum through their $x$ dependence. For the gluons our
transformation changes $\bn\cdot A_n$ but not the $n\cdot A_n$ field, whereas
Eq.~(\ref{BF}) does the exact opposite.  For the $A_n^\perp$ field our hatted
field is not surrounded by W's, and we have a covariant usoft derivative while
Eq.~(\ref{BF}) has a normal derivative.  The fact that both our usoft and
collinear fields are local in the coordinate $x$ representing residual momenta
$k^\mu\sim Q\lambda^2$ means that we did not need to consider a matrix like $R$.
Also, note that in our procedure for transforming the fields we did not require
any gauge fixing at intermediate steps. Finally, we comment that the form of our
field redefinition leads to an interesting result for ${\cal W}$ in terms of the
new fields, namely ${\cal W}=\hat W$ with no higher order terms in $\lambda$.

The use of position and momentum space makes a more direct comparison difficult.
However, any field redefinitions that lead to the desired result are equally
valid and both Eq.~(\ref{BF}) and Eq.~(\ref{redef2}) satisfy this criteria. In
general one knows that field redefinitions should only affect the form of
operators and the result for Green's functions, but should not affect S-matrix
elements. Thus, equivalent effective theories are often realized with different
fields. We expect that there should be a field redefinition which would relate
our fields $\hat A_n$ to the fields $\hat A_n$ in Ref.~\cite{bf2}, although we
have not constructed it in closed form.

\vspace{0.2cm}
\noindent \underline{\bf Lagrangian Results}\\[-5pt]

Having established collinear gauge fields whose transformations never mix orders
in $\lambda$, we now rewrite all subleading Lagrangians to order $\lambda^2$
using Eq.~(\ref{redef2}).  For simplicity we omit the hats in the following
equations, however all collinear gauge fields should be understood to be the
hatted ones. For the collinear quark Lagrangian we find
\begin{eqnarray} \label{Lxxnew}
{\cal L}_{\xi\xi}^{(1)} 
  &=&  \big(\bar \xi_n  W\big)\,  i\Dslash^\perp_{us} \frac{1}{\bnP} \big(W^\dagger 
  i \Dslash^\perp_c  \frac{\bnslash}{2} \xi_n \big)
  + \big( \bar \xi_n  i \Dslash^{\perp}_c W\big) \frac{1}{\bnP} i\Dslash^\perp_{us}
   \big( W^\dagger \frac{\bnslash}{2} \xi_n \big)\nn\\
  {\cal L}_{\xi\xi}^{(2)} &=&  \big( \bar \xi_n W\big)
 i\Dslash_{us}^\perp \frac{1}{\bnP} i\Dslash_{us}^\perp \frac{\bnslash}{2}
  \big( W^\dagger \xi_n\big) +
 \big( \bar \xi_n i\Dslash^\perp_c W\big) \frac{1}{\bnP^2} i\bn\mcdot D_{us} 
  \frac{\bnslash}{2}\big( W^\dagger i\Dslash^\perp_c \xi_n\big) \,, 
\end{eqnarray} 
where we have used the fact that
\begin{eqnarray}
  \frac{1}{i\bn\mcdot D} = \frac{1}{i\bn\mcdot D_c} 
   - W \frac{1}{\bnP^2} i\bn\mcdot D_{us} W^\dagger + \ldots \,.
\end{eqnarray}
It is easy to see that the results in Eq.~(\ref{Lxxnew}) are invariant under the
transformations in Table~\ref{table_gt}.
For the mixed collinear-usoft quark interactions we find the invariant results
\begin{eqnarray} \label{Lxiq}
  {\cal L}^{(1)}_{\xi q} &=&   \bar\xi_n \: \frac{1}{i\bn\mcdot D_c}\: 
 ig\, \Bslash_\perp^c W  q_{us} \mbox{ + h.c.}\,,\nn\\
    {\cal L}^{(2a )}_{\xi q} &=& \bar\xi_n \frac{\bnslash}{2}
     \frac{1}{i\bn\mcdot D_c}\: 
     ig\,  n\mcdot M \, W \, q_{us} \mbox{ + h.c.} \,, \nn\\
  {\cal L}^{(2b)}_{\xi q} &=& \bar\xi_n \frac{\bnslash}{2} 
  i\Dslash_\perp^{\,c} \frac{1}{(i\bn\mcdot D_c)^2}\:   ig\, \Bslash_\perp^c W 
  \: q_{us}   \mbox{ + h.c.}\,, 
\end{eqnarray}
where $ig \Bslash_\perp^c =[i\bn\mcdot D^c,i\Dslash_\perp^{c}]$ and we have used
the fact that the transformation of ${\cal L}_{\xi q}^{(1)}$ makes
\begin{eqnarray}
  ig M_\perp^\mu &=& [i\bn\mcdot D_c , W iD_{us\perp}^\mu W^\dagger ] 
  = [ W \bnP W^\dagger, W i D_{us\perp}^\mu W^\dagger ] 
  = W [\bnP , i D_{us\perp}^\mu ] W^\dagger  =  0 \,.
\end{eqnarray}
Finally, for the subleading terms in the mixed usoft-collinear gluon action we find
\begin{eqnarray}
 {\cal L}_{cg}^{(1)} &=& \frac{2}{g^2}\: {\rm tr} 
  \Big\{ \big[i {D}_0^\mu , iD_c^{\perp\nu} \big] 
         \big[i {D}_{0\mu} , W iD_{us\,\nu}^\perp W^\dagger \big] \Big\}\,,\\
 {\cal L}_{cg}^{(2)} &=&  \frac{1}{g^2}\: {\rm tr} 
  \Big\{ \big[i {D}_0^\mu , W iD_{us}^{\perp\nu} W^\dagger \big] 
         \big[i {D}_{0\mu} , W iD_{us\,\nu}^\perp W^\dagger \big] \Big\} \nn\\
 &&\!\!\!\!\!\!\!\!\!\!\! 
  +\frac{1}{g^2}\: {\rm tr} 
  \Big\{ W \big[ i D_{us}^{\perp\mu} , iD_{us}^{\perp\nu} \big] W^\dagger 
         \big[i {D}_{c\mu}^\perp , i{D}_{c\nu}^\perp \big] \Big\} 
 +\frac{1}{g^2}\: {\rm tr} 
  \Big\{ \big[i {D}_0^\mu , i n\mcdot D \big] 
         \big[i {D}_{0\mu} , W i \bn\mcdot D_{us} W^\dagger \big] \Big\} \nn \\
 &&\!\!\!\!\!\!\!\!\!\!\!
   + \frac{1}{g^2}\: {\rm tr} 
  \Big\{ \big[W i D_{us}^{\perp\mu} W^\dagger , iD_{c}^{\perp\nu} \big] 
         \big[i {D}_{c\mu}^\perp , W i{D}_{us\nu}^\perp W^\dagger \big]
         \Big\}\nn \,,
\end{eqnarray}
where $i D_0^\mu = i{\cal D}^\mu + g A_{n}^\mu$.

\section{Power Suppressed Soft-Collinear Operators}\label{sect2}

In SCET$_{\rm II}$ the structure of operators with soft and collinear fields is
still constrained by properties such as power counting, gauge invariance, and
reparameterization invariance. However the non-local nature of the theory makes
it more difficult to simply write down the most general operators in an
arbitrary case.  To see this we consider a simple example, namely a
heavy-to-light current. In the full theory we have $\bar q \Gamma b$ and in the
effective theory
\begin{eqnarray}\label{heavylight}
 C(\bnP)\: \bar \xi_n W \Gamma S^\dagger h_v. 
\end{eqnarray}
The Wilson lines $W$ and $S$ are required to ensure collinear and soft gauge
invariance respectively.  However, neither gauge invariance nor power counting
determines the exact path of $S$ from $x$ to $\infty$, since all $A_s^\mu$
fields scale the same way.  Thus, additional input is needed to constrain these
operators.  From direct matching calculations, which integrate out fluctuations
with $p^2\sim Q\Lambda$, it is straightforward to determine that $S$ is a
straight Wilson line along the $n$ direction built out of $n\mcdot A_s$
fields~\cite{bps2}. An alternative procedure is~\cite{bps4}
\begin{eqnarray}
  && \mbox{ i)\phantom{ii}\quad Match QCD onto SCET$_{\rm I}$ at a scale 
    $\mu^2\sim Q^2$ (with $p_c^2\sim Q\Lambda$)}\nn \\
  && \mbox{ ii)\phantom{i}\quad Factorize the usoft-collinear interactions 
     with the field redefinitions, } \nn\\ 
  && \mbox{\hspace{2cm}
     $\xi_n = Y\xi_n^{(0)}$ and $A_n^\mu = Y A_n^{(0)\mu} Y^\dagger $.}\nn \\
  && \mbox{ iii)\quad  Match SCET$_{\rm I}$ onto SCET$_{\rm II}$ at a scale 
    $\mu^2\sim Q\Lambda$ (with $p_c^2\sim \Lambda^2$) } \,. \nn \qquad\qquad\qquad
\end{eqnarray} 
For the heavy-to-light case we have i) $\bar q \Gamma b \to C(\bnP)\: \bar \xi_n
W \Gamma\: h_v^{us}$, and then ii) $C(\bnP)\: \bar \xi_n W \Gamma\: h_v^{us} =
C(\bnP)\: \bar \xi_n^{(0)} W^{(0)} \Gamma\: Y^\dagger h_v^{us}$. For the final
step we rename the usoft fields as soft fields $Y^\dagger h_v^{us}=S^\dagger
h_v^{s}$, and then lower the offshellness of the collinear fields. Since the
leading collinear Lagrangians in \SCETa and \SCETb are the same all possible
time-ordered products agree exactly and we can simply replace $C(\bnP)\: \bar
\xi_n^{(0)} W^{(0)}\to C(\bnP)\: \bar \xi_n^{\rm II} W^{\rm II}$. The final
result is identical to Eq.~(\ref{heavylight}) but the steps are simpler than
those carried out in the appendix of Ref.~\cite{bps2}. From the two-step
approach it is also clear why the Wilson coefficient does not pick up any
dependence on the soft momentum in this example.

The two-stage matching procedure becomes even more useful in cases where \SCETa
contains time-ordered products, since these can induce non-trivial jet functions
involving $p^2\sim Q\Lambda$ fluctuations.  SCET$_{\rm I}$ gives a well defined
set of Feynman rules for computing these jet functions at tree level and in
loops, and does so in a manner independent from the computation of Wilson
coefficients at the hard scale $p^2\sim Q^2$. Since the operator in \SCETa is a
time-ordered product we are guaranteed that the running to the scale
$\mu^2=Q\Lambda$ is determined by that of the product of the hard Wilson
coefficients.  A final benefit is that power counting in \SCETa
constrains the allowed scaling of operators in \SCETb, and in particular, places
a limit on the number of factors of $1/\Lambda$ that can be induced from
$1/(Q\Lambda)$ terms as we discuss below.  This provides a complementary
procedure to constraining the powers of $1/\Lambda$ with reparameterization
invariance as first described in Ref.~\cite{HN}.

Let us consider a generic matching calculation 
\begin{eqnarray} \label{M12}
  {\rm SCET}_{\rm I}\ [ p_c^2 \sim Q\Lambda\,, p_{us}^2\sim \Lambda^2]
   \ \ \stackrel{\mu^2\sim Q\Lambda}{\longrightarrow} \ \ 
  {\rm SCET}_{\rm II}\ [ p_c^2 \sim \Lambda^2\,,  p_s^2\sim\Lambda^2] \,.
\end{eqnarray}
First construct all time-ordered products, $T_{\rm I}^j$, of SCET$_{\rm I}$
operators which contribute at a given order in the power counting. To match
these onto SCET$_{\rm II}$ operators we take matrix elements,
\begin{eqnarray} \label{melt}
 \langle\: \phi_I(p_i^2\sim \Lambda^2)\: | \: T^j_I\:
  |\: \phi'_I(p_i^2\sim \Lambda^2)\: \rangle\,.
\end{eqnarray} 
Here the states have particles with ultrasoft momenta $p_{us}^2\sim \Lambda^2$,
but with small collinear momenta $p_c^2\sim \Lambda^2$. These are allowed states
in the Hilbert space of SCET$_{\rm I}$, since for example $p_\perp^2$ momenta of
this size correspond to having zero label $\perp$ momenta, but non-zero residual
$\perp$ momenta.  These are also obviously states in SCET$_{\rm II}$.  As in any
matching calculation, we can use any convenient states, and one usually chooses
free particle states.  Note that the external collinear particles in (\ref{melt})
have reduced offshellness, however this is not in general the case for the internal
propagators.  

As an additional constraint, the matching in Eq.~(\ref{M12}) must be carried out
in a manner that accounts for the fact that only certain products of collinear
fields have {\em gauge invariant} label momentum, and that these momentum
components are not lowered in matching these products of fields onto collinear
fields in \SCETb.  This means that only gauge invariant products of collinear
fields should be integrated out in the matching (guaranteeing that gauge
invariant products are also left over). This automatically builds in the fact
that the low energy operators in \SCETb must be built out of gauge invariant
products $\Phi_1 = W^\dagger \xi_n$, $\Phi_2 = [W^\dagger D_c^\perp W]$, ${\cal
  S}_1 = S^\dagger q_s$, etc.  This properly matches the theory \SCETa onto the
subset of phase space that is described by fields in \SCETb. This matching will be
perturbative as long as the scale $\sqrt{Q\Lambda}\gg\Lambda $.

A useful benefit of the two-stage procedure is that the power counting is
transparent. Thus even though we are integrating out an intermediate scale
$p^2\sim Q\Lambda$ that involves factors of the hadronic scale $\Lambda$, we need
not worry about missing operators that would be power suppressed but are
enhanced by explicit factors of $1/\Lambda$.  The power counting for the
matching process is
\begin{eqnarray} \label{pc12}
  T^I \sim \lambda^{2k}  \rightarrow {\cal O}^{II} \sim \eta^{k+E} \,,
\end{eqnarray}
where the final scaling is independent of how factors of $\eta$ are partitioned
between coefficients and operators in SCET$_{\rm II}$ (we will choose to make
Wilson coefficients in SCET$_{\rm II}$ dimensionless and order $\eta^0$).  This
equation says that T-products which are order $\lambda^{2k}$ in SCET$_{\rm I}$
will match onto operators in SCET$_{\rm II}$ that are order $\eta^{k+E}$ with
$E\ge 0$. Here the factor $\eta^E$ is the extra factor obtained by lowering the
offshellness of the external collinear fields and thereby changing their power
counting.  For example $(\xi_n^{\rm I} \sim\lambda=\sqrt{\eta}) \to (\xi_n^{\rm
  II}\sim \eta)$, which agrees with the formula having $E=1/2$.  In general
$E=1/2$ for external $\xi_n$ or $A_n^\perp$, $E=0$ for external $\bn\mcdot A_n$
or $W$, and $E=1$ for external $n\mcdot A_n$.

To illustrate these points we consider several examples. First consider the
example of factorization in $B\to D\pi$~\cite{bps}, but using the two-stage
procedure.  Matching the two $(\bar c b)_{V-A} (\bar d u)_{V-A}$ electroweak
four quark operators onto operators in SCET$_{\rm I}$ gives
\begin{eqnarray}
 {Q}_{\bf 0}^{\rm I} &=& \big[{\bar  h_{v'}^{us}} \, \,\Gamma_h
    {h_v^{us}}\big] \big[{ \bar\xi_{n,p'} W} 
   {C_{\bf 0}(\bnP_+)} \Gamma_l { W^\dagger  \xi_{n,p} }\big] \,,\\
 {Q}_{\bf 8}^{\rm I} &=& \big[{ \bar h_{v'}^{us} } \, \Gamma_h T^A {
    h_v^{us}}\big] \big[{ \bar\xi_{n,p'} W} { C_{\bf 8}(\bnP_+)}\Gamma_l
       T^A { W^\dagger  \xi_{n,p} }\big] \nonumber \,,
\end{eqnarray}
where $\bnP_+ = \bnP^\dagger + \bnP$ and the Wilson coefficients $C_{\bf 0,8}$
contain the hard $p^2\sim Q^2$ effects. Next decouple the usoft interactions
from the leading collinear Lagrangian with the field redefinitions $\xi_n =
Y\xi_n^{(0)}$ and $A_n^\mu=  Y A_n^{(0)\mu} Y^\dagger$~\cite{bps2}. This leaves
\begin{eqnarray}
 {Q}_{\bf 0}^{\rm I} &=& \big[{\bar  h_{v'}^{us}} \,\Gamma_h \,
    {h_v^{us}}\big] \big[{ \bar\xi_{n,p'}^{(0)} W^{(0)} } 
   {C_{\bf 0}(\bnP_+)}\Gamma_l { W^{(0)\dagger}  \xi_{n,p}^{(0)} }\big] \,, \\
 {Q}_{\bf 8}^{\rm I} &=& \big[{ \bar h_{v'}^{us} } \,\Gamma_h Y T^A Y^\dagger {
    h_v^{us}}\big] \big[{ \bar\xi_{n,p'}^{(0)} W^{(0)} } { C_{\bf 8}(\bnP_+)}
   \Gamma_l
       T^A { W^{(0)\dagger}  \xi_{n,p}^{(0)} }\big] \nonumber \,.
\end{eqnarray}
In this result the ultrasoft and collinear fields are completely factorized. The
collinear fields still have large offshellness $p^2\sim Q\Lambda$, so we need
step iii).  Taken with leading order Lagrangian insertions this example is
just like the heavy-to-light current, so we match directly onto the \SCETb
operators
\begin{eqnarray}
 {Q}_{\bf 0}^{\rm II} &=& \big[{\bar  h_{v'}^{s}} \, \Gamma_h \,
    {h_v^{s}}\big] \big[{ \bar\xi_{n,p'} W} 
   {C_{\bf 0}(\bnP_+)}\Gamma_l { W^\dagger  \xi_{n,p} }\big] \,,\\
 {Q}_{\bf 8}^{\rm II} &=& \big[{ \bar h_{v'}^{s} } \, \Gamma_h S T^A S^\dagger {
    h_v^{s}}\big] \big[{ \bar\xi_{n,p'} W} { C_{\bf 8}(\bnP_+)}\Gamma_l
       T^A { W^\dagger  \xi_{n,p} }\big] \nonumber \,.
\end{eqnarray}
This is the same as the result originally derived in Ref.~\cite{bps}. It is easy
to see that no other \SCETb operators are possible at this order.

This algebra was quite simple, however we have not yet seen the full power of the
intermediate theory with the above example. The procedure becomes useful
once we consider time-ordered products in \SCETa, since then one can obtain
non-trivial jet functions $J$ in \SCETb which lead to Wilson coefficients
$C(z_i)\: J(z_i,x_j,y_k)$ for the SCET$_{\rm II}$ operators. This jet function
has convolutions with variables $z_i$ that correspond to the $p^-$ momentum
dependence in the hard coefficient $C$. It also can have dependence on the $x_j$
momentum fractions of collinear fields in the \SCETb operators we match onto.
Finally, since collinear fields in \SCETa are affected by the $k^+$ usoft
momenta (through the $in\mcdot \partial$ term in their action) the jet $J$ can
depend on the momentum fractions $y_k$ which correspond to the soft $+$-momenta of
gauge invariant products of soft fields in \SCETb.

An example of a more involved matching calculation was given for the case of
heavy-to-light form factors in Refs.~\cite{bps4,ps1} and we will not repeat this
example here. To illustrate this case of matching further consider the toy
example of light-light soft-collinear currents. In Ref.~\cite{HN} these currents
were derived by direct matching from QCD, so we contrast this procedure with the
matching onto \SCETb operators by using \SCETa\!.  Such operators are matched
from contributions in \SCETa which provide mixing between collinear and usoft
quarks. Consider
\begin{eqnarray} \label{Ts}
  T_0^{(3)} &=& \int\! d^4 x\, T[ J_{\xi\xi}^{(2)}(0) , i {\cal L}_{\xi
    q}^{(1)}(x) ]
 \nn\\
  J_{\xi q}^{(4)} &=& \bar\xi_n W \Gamma q_{us}
  \,,
\end{eqnarray}
where $J_{\xi\xi}^{(2)} = \bar\xi_n W \Gamma W^\dagger \xi_n$ and ${\cal L}_{\xi
  q}^{(1)}(x)$ is given in Eq.~(\ref{Lxiq}) (hard coefficients are suppressed
since they are not crucial to our discussion). The order in $\lambda$ is denoted
by the exponent in brackets. To match these operators onto \SCETb we use the
procedure explained above. For the local operator $J_{\xi q}^{(4)}$ this
matching is simple. We first perform the field redefinition $\xi_n =
Y\xi_n^{(0)}$ and $A_n^\mu = Y A_n^{(0)\mu} Y^\dagger$ to write
\begin{eqnarray}
  J_{\xi q}^{(4)} &=& \left[ \bar\xi^{(0)} _n W^{(0)} \right] \Gamma 
 \left[Y^\dagger q_{us}\right]
\end{eqnarray}
where we have indicated the gauge invariant blocks of fields by the square
brackets. The final step is to identify the usoft fields with soft fields and to
lower the off-shellness of the collinear fields. At tree level this leads to the
operator
\begin{eqnarray} \label{O1}
 O_1 = [\bar\xi_n W] \Gamma [ S^\dagger q_s ]
\end{eqnarray}
in \SCETb which is order $\eta^{5/2}$. This follows from Eq.~(\ref{pc12}) with
$k=2$ and $E=1/2$.

\begin{figure}[!t]
\vskip-0.3cm
 \centerline{
  \mbox{\epsfxsize=5truecm \hbox{\epsfbox{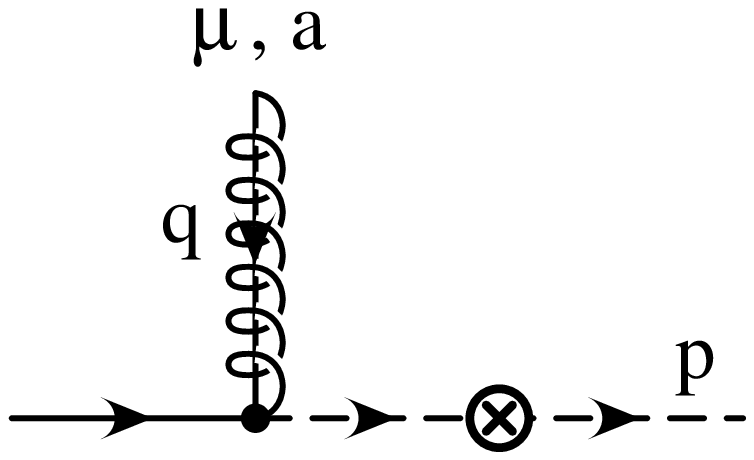}} }\hspace{1.4cm}
  \mbox{\epsfxsize=5.5truecm \hbox{\epsfbox{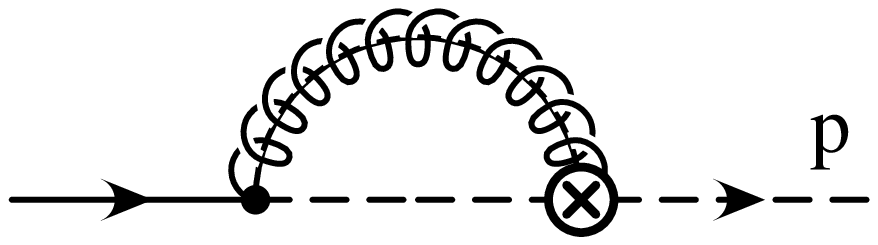}} }
  } 
\vskip-0.3cm
\caption[1]{Examples of graphs contributing to the matching of the SCET$_{\rm I}$
  T-products onto SCET$_{\rm II}$ operators in
  Eq.~(\ref{O12s}). The dots denote the insertion of a
  ${\cal L}_{\xi q}^{(1)}$ and the circled crosses in the two diagrams
  are $J_{\xi\xi}^{(2,3)}$ operators respectively.}
\label{fig_match} 
\vskip-0.5cm
\end{figure}

For the time-ordered product $T_0^{(3)}$ we follow similar steps. After the
field redefinition 
\begin{eqnarray}
T_0^{(3)} \!\!&=&\!\! \int\!\! d^4x\, T\Big\{ \left[\bar\xi_n^{(0)}
    W^{(0)}\right] \Gamma 
 \left[W^{(0)\dagger} \xi_n^{(0)}\right](0) , \left[\bar\xi_n^{(0)} W^{(0)}
 \right] \left[ W^{(0)\dagger} i\Dslash_\perp^c W^{(0)} \right] \left[Y^\dagger
   q_{us} \right](x) \Big\}.\ \ \
\end{eqnarray}
Consider the matrix element between a collinear fermion, a $\perp$ collinear
gluon and a soft fermion. To match onto \SCETb we contract the $[W^{(0)\dagger}
\xi_n^{(0)}][\bar\xi_n^{(0)} W^{(0)}]$ product, lower the off-shellness of the
remaining $[\bar\xi_n^{(0)} W^{(0)}]$ and $[ W^{(0)\dagger} i\Dslash_\perp^c
W^{(0)}]$ and rename the $[Y^\dagger q_{us}]$ to $[S^\dagger q_{s}]$.  At tree
the two collinear fermion fields get contracted giving a propagator as shown in the
first diagram of Fig.~\ref{fig_match}.  This gives the operator
\begin{eqnarray} \label{O2}
O_2 = [\bar\xi_n W] \Gamma \frac{\nslash}{2}
  [W^\dagger i\Dslash_\perp^c W ] \frac{1}{n\mcdot \cP} [S^\dagger q_s]
\end{eqnarray}
in \SCETb which is the same operator as Ref.~\cite{HN}. Note that while in
\SCETa $T_{0}^{(3)}$ was larger by one power of $\lambda$ than $J_{\xi
  q}^{(4)}$, the resulting two operators are the same order in $\eta$. This is
because in lowering the off-shellness of $[ W^{(0)\dagger} i\Dslash_\perp^c
W^{(0)}]$ the power counting of the $\perp$ gluon is reduced from $\lambda$ to
$\eta=\lambda^2$. This agrees with Eq.~(\ref{pc12}) with $E=1/2$, so $O_2\sim
\eta^{5/2}$ just like $O_1$.\footnote{Note that in matching we always expand the
  upper theory in a series of terms to match it onto the lower
  theory. Therefore, it is not unusual that operators in \SCETa match onto  
  operators of different orders in \SCETb.}

There are additional contributions in \SCETa that one can write down at order
$\lambda^4$, such as $T[ J_{\xi\xi}^{(2)}(0), i {\cal L}_{\xi q}^{(2)}(x) ]$,
$T[ J_{\xi\xi}^{(2)}(0), i {\cal L}_{\xi q}^{(1)}(x),i {\cal L}_{\xi q}^{(1)}(y)
] $, and $T[ J_{\xi\xi}^{(3)}(0) , i {\cal L}_{\xi q}^{(1)}(x) ]$. At tree level
all these contributions contain factors of $D^c$, which receive an additional
suppression factor when matching onto \SCETb.  However, at higher orders in
perturbation theory these operators can contribute since more collinear fields
are contracted.  For the operators displayed in Eqs.~(\ref{O1},\ref{O2}) they
give rise to non-trivial jet functions. Consider for example the time-ordered
product
\begin{eqnarray}
  T_0^{(4)} = \int\!\! d^4x\ T[ J_{\xi\xi}^{(3)}(0), i {\cal L}_{\xi q}^{(1)}(x) ]
\end{eqnarray}
where $J_{\xi\xi}^{(3)}= (\bar\xi_n W) \Gamma (1/\bnP) (W^\dagger
i\Dslash_\perp^{\:c} W) (\bnslash/2) (W^\dagger \xi_n)$.  Operators like
$T_0^{(4)}$ appear for example in the matching of QCD onto \SCETa for the
electromagnetic current of light quarks (see the second reference
in~\cite{bgamenu}). Gauge invariant blocks of collinear fields in the
time-ordered product are contracted when matching onto \SCETb.  An example is
illustrated in the second diagram in Fig.~\ref{fig_match} where the factors of
fields containing $D_\perp^c$ derivatives are contracted.  Such a graph does not
exhibit the additional suppression factor, as there is no collinear covariant
perpendicular derivative left over.  Thus, this operator can contribute to the
operator $O_1$ and induce a non-trivial Wilson coefficient $J$. Therefore, the
operators ${\cal O}_{1,2}$ in \SCETb contributing to light-light soft-collinear
current at any order in the matching from \SCETa have the form
\begin{eqnarray} \label{O12s}
 O_1 &=& J_1(\omega,y)\, (\bar\xi_n W)_\omega \Gamma (S^\dagger q_s)_y \,,\nn\\
 O_2 &=& J_2(\omega_i,y)\, (\bar\xi_n W)_{\omega_1} \Gamma \frac{\nslash}{2}
  [W^\dagger i\Dslash_\perp^c W ]_{\omega_2} \frac{1}{n\mcdot \cP} 
  (S^\dagger q_s)_y \,,
\end{eqnarray}
where $(\bar\xi_n W)_\omega = [\bar\xi_n W \delta(\omega-\bnP^\dagger)]$ and
$(S^\dagger q_s)_y = [\delta(y-n\mcdot P) S^\dagger q_s]$.

Finally, this procedure can also be used to match onto the Lagrangian for mixed
soft-collinear interactions in SCET$_{\rm II}$.  After making the field
redefinition in step ii) there are no usoft-collinear Lagrangian interactions at
order $\lambda^0$ in \SCETa. Therefore from Eq.~(\ref{pc12}) it follows that it
is not possible to construct a gauge invariant order $\eta^0$ soft-collinear
Lagrangian. This is true for both quarks and gluons. This very simple power
counting argument clarifies the original argument based on gauge invariance and
power counting in Ref.~\cite{bps2} and supplements the direct matching
calculations in Ref.~\cite{HN}. In the language of the power counting formulae in
Ref.~\cite{bpspc} the power counting for soft-collinear Lagrangian terms in
\SCETb corresponds to an index factor $(k-3) V_{SC}^k$ in the equation for
$\delta$ which gives the power counting for an arbitrary time-ordered product.
Here $V_{SC}^k$ counts the number of insertions of soft-collinear Lagrangian
operators that are order $\eta^k$.  The factor of $(k-3)$ agrees with the phase
space argument in Ref.~\cite{HN}.

At order $\lambda^2$ we have a time-ordered product $\int d^4x T\{ {\cal L}_{\xi
  q}^{(1)}(0), i{\cal L}_{\xi q}^{(1)}(x) \}$, which can induce suppressed
operators in the \SCETb Lagrangian. Contracting the collinear quarks in a
$W^\dagger \xi_n(0) \bar\xi_n W(x)$ factor this gives an operator whose form
agrees with Eq.~(17) of Ref.~\cite{HN}. At tree level in the matching we find
\begin{eqnarray} \label{LS1}
  {\cal L}_{qqBB}^{(1)} &=& (\bar q_s S) \Big( W^\dagger i g \Bslash_\perp^c W
  \frac{1}{\bnP^\dagger} \Big) \frac{\nslash}{2} \Big( \frac{1}{\bnP} W^\dagger
  i g \Bslash_\perp^c W \Big) \frac{1}{n\mcdot \cP} ( S^\dagger q_s ) \,.
\end{eqnarray}
Here the factor $\nslash/(2 n\mcdot \cP)$ is again from the collinear quark
propagator, and from Eq.~(\ref{pc12}) we count $E=1$ since two $\perp$ gluons
are external and have their power counting changed in passing to \SCETb.  The
superscript $(1)$ indicates that this operator contributes at order $\eta$ in
\SCETb.  The factor of $\eta$ is derived by noting that the operator in
Eq.~(\ref{LS1}) is $\sim \eta^4$ and so counts as $V_{SC}^4=1$. Thus subtracting
three we see that it contributes an $\eta$ to the $\delta$ power counting
formula.

\section{Conclusion}

In this paper we discussed a few issues related to the gauge invariance of the
soft-collinear effective theory beyond leading order. Together with power
counting and reparameterization invariance, gauge invariance constrains the form
of the allowed effective theory operators. However, there is some freedom in 
splitting the QCD gluon field into collinear and ultrasoft fields in the
effective theory. In Sec.~II we showed that the choice which gives
\begin{eqnarray} \label{final}
  && \qquad \qquad\quad
  i\bn\mcdot \hat D = i\bn\mcdot \hat D_c + {\hat W} i\bn\mcdot D_{us}
   {\hat W}^\dagger  \,,\qquad
  i \hat D_\perp^\mu = i \hat D_{c,\perp}^\mu + {\hat W} iD_{us,\perp}^\mu
   {\hat W}^\dagger  \,,\qquad\qquad
\end{eqnarray}
corresponds to collinear and usoft fields which transform in a homogeneous way
under the gauge transformations at any order in $\lambda$.  This result
uniquely fixes how power suppressed ultrasoft derivatives appear which are
related to the collinear derivatives by reparameterization invariance.  Using
the new fields we then gave results for the subleading collinear and
usoft-collinear effective Lagrangians to $O(\lambda^2)$, which by themselves are
invariant under the collinear gauge transformations in Table~\ref{table_gt}.

A related construction was presented in Ref.~\cite{bf2} using a position space
multipole expansion. The collinear field redefinition adopted here differs from
the one there. Our construction has the benefit of avoiding gauge fixing in the
derivation. The explicit form of the transformation relating the fields in
Ref.~\cite{bf2} to the fields we have here remains an open and interesting
question.

For SCET$_{\rm II}$, power counting, RPI and gauge invariance also give
restrictions on allowed operators, which are however not as strict as in
SCET$_{\rm I}$. The reason is that \SCETb is non-local at the scale over which
soft particles are propagating, whereas \SCETa is only non-local at the hard
scale $Q$. (This is the case before we decide to induce by hand a non-local $Y$
in \SCETa by making a field redefinition.) Thus, additional input is needed to
construct operators in SCET$_{\rm II}$, and one has to carefully consider which
modes are integrated out in arriving at the low energy theory. In
Ref.~\cite{bps4} it was proposed that soft-collinear operators in SCET$_{\rm
  II}$ could be constructed in an elegant manner by making use of factored
ultrasoft-collinear operators in SCET$_{\rm I}$. In Sec.~III we presented
details of this matching calculation in several examples, and showed how the
constraints from power counting and gauge invariance on SCET$_{\rm I}$ restrict
the form of the operators induced in matching onto SCET$_{\rm II}$.

\acknowledgments{
  This work was supported
  in part by the U.S. Department of Energy (DOE) under the cooperative research 
  agreement DF-FC02-94ER40818, the grant DOE-FG03-97ER40546, as well as the NSF 
  under grant PHY-9970781. }
 


\begin{thebibliography}{99}

\bibitem{bfl}
C.~W.~Bauer, S.~Fleming and M.~Luke,
Phys.\ Rev.\ D {\bf 63}, 014006 (2001) [arXiv:hep-ph/0005275].
%
\bibitem{bfps}
C.~W.~Bauer, S.~Fleming, D.~Pirjol and I.~W.~Stewart,
Phys.\ Rev.\ D {\bf 63}, 114020 (2001) [arXiv:hep-ph/0011336].

\bibitem{cbis}
C.~W.~Bauer and I.~W.~Stewart,
Phys.\ Lett.\ B {\bf 516}, 134 (2001).
[arXiv:hep-ph/0107001].

\bibitem{bps2}
C.~W.~Bauer, D.~Pirjol and I.~W.~Stewart,
Phys.\ Rev.\ D {\bf 65}, 054022 (2002) [arXiv:hep-ph/0109045].

\bibitem{reviews}
For reviews R.~L.~Jaffe,
[arXiv:hep-ph/9602236];
G. Sterman, TASI lectures 1995, [arXiv:hep-ph/9606312];
J.C.~Collins, D.E.~Soper, and G.~Sterman in {\em Perturbative
Quantum Chromodynamics}, Ed. by A. H. Mueller, World Scientific
Publ., 1989, p. 1-93.
S.~J. Brodsky and G.~P. Lepage, in {\em Perturbative Quantum 
Chromodynamics}, pp. 93-240; S.~J. Brodsky, SLAC report SLAC-PUB-9281, 
arXiv:hep-ph/0208158; 
J.~C.~Collins and D.~E.~Soper,
Ann.\ Rev.\ Nucl.\ Part.\ Sci.\  {\bf 37}, 383 (1987).

\bibitem{bfprs}
C.~W.~Bauer, S.~Fleming, D.~Pirjol, I.~Z.~Rothstein, I.~W.~Stewart,
Phys.\ Rev.\ D {\bf 66}, 014017 (2002) [arXiv:hep-ph/0202088].

\bibitem{jets}
C.~W.~Bauer, A.~V.~Manohar and M.~B.~Wise,
arXiv:hep-ph/0212255.

\bibitem{bps4}
C.~W.~Bauer, D.~Pirjol and I.~W.~Stewart,
arXiv:hep-ph/0211069.

\bibitem{bgamenu}
S.~Descotes-Genon and C.~T.~Sachrajda,
Nucl.\ Phys.\ B {\bf 650}, 356 (2003) 
[arXiv:hep-ph/0209216];
E.~Lunghi, D.~Pirjol and D.~Wyler,
Nucl.\ Phys.\ B {\bf 649}, 349 (2003)
[arXiv:hep-ph/0210091]; 
see also S.~W.~Bosch, R.~J.~Hill, B.~O.~Lange and M.~Neubert,
arXiv:hep-ph/0301123.

\bibitem{HN}
R.~J.~Hill and M.~Neubert,
arXiv:hep-ph/0211018.

\bibitem{chay}
J.~Chay and C.~Kim,
Phys.\ Rev.\ D {\bf 65}, 114016 (2002)
[arXiv:hep-ph/0201197].

\bibitem{chay2}
J.~Chay and C.~Kim,
arXiv:hep-ph/0205117.

\bibitem{bpspc}
C.~W.~Bauer, D.~Pirjol and I.~W.~Stewart,
Phys.\ Rev.\ D {\bf 66}, 054005 (2002)
[arXiv:hep-ph/0205289].

\bibitem{bcdf}
M.~Beneke, A.~Chapovsky, M.~Diehl, T.~Feldmann, 
Nucl.\ Phys.\ B {\bf 643}, 431 (2002) [arXiv:hep-ph/0206152].

\bibitem{bf2}
M.~Beneke and T.~Feldmann,
Phys.\ Lett.\ B {\bf 553}, 267 (2003) [arXiv:hep-ph/0211358].

\bibitem{ps1}
D.~Pirjol and I.~W.~Stewart,
arXiv:hep-ph/0211251.

\bibitem{ira}
I.~Z.~Rothstein,
arXiv:hep-ph/0301240.

\bibitem{adam}
A.~K.~Leibovich, Z.~Ligeti and M.~B.~Wise,
arXiv:hep-ph/0303099.

\bibitem{rpi}
A.~V.~Manohar, T.~Mehen, D.~Pirjol, I.~W.~Stewart,
Phys.\ Lett.\ B {\bf 539}, 59 (2002)
[arXiv:hep-ph/0204229].

\bibitem{rev2}
I.~W.~Stewart,
Nucl.\ Phys.\ B (Proc.\ Suppl.) {\bf 115}, 107 (2003) 
[arXiv:hep-ph/0209159].

\bibitem{LM}
M.~E.~ Luke and A.~V.~ Manohar,
Phys.\ Lett.\ B {\bf 286}, 348 (1992) [arXiv:hep-ph/9205228].

\bibitem{feldmann}
T.~Feldmann,
arXiv:hep-ph/0209239.

\bibitem{bps}
C.~W.~Bauer, D.~Pirjol and I.~W.~Stewart,
Phys.\ Rev.\ Lett.\  {\bf 87}, 201806 (2001)
[arXiv:hep-ph/0107002].


\end{thebibliography}
\end{document}